\documentclass{svjour3}
\usepackage{times} 
\usepackage{cite} 
\usepackage{hyperref}
\usepackage{blindtext, graphicx}
\usepackage{cite}
\usepackage{url}
\usepackage{multirow} 
\usepackage{tabu}
\usepackage{colortbl}
\usepackage{wrapfig}
\usepackage{tikz}
\usepackage{wrapfig}
\usepackage{adjustbox}
\usepackage{fixltx2e}
\usepackage{hyperref}
\usepackage{amssymb}
\usepackage{pifont}
\usepackage{tcolorbox}

\usepackage{tikz}

\usepackage{graphicx}

\smartqed
\newcommand{\bi}{\begin{itemize}}

\newcommand{\ei}{\end{itemize}}

\usepackage[tikz]{bclogo}
{\noindent\begin{minipage}[c]{\linewidth}%
\begin{bclogo}[couleur=gray!25,%
arrondi=0.1,%
logo=\bctrombone,%
ombre=true]{{\small ~#1}}}%
{\end{bclogo}\end{minipage}\vspace{2mm}}

\newcommand{\tbl}[1]{Table~\ref{tbl:#1}}
\newcommand{\fig}[1]{Figure~\ref{fig:#1}}
\usepackage{balance}
\usepackage{pifont}
\definecolor{Gray}{rgb}{0.88,1,1}
\definecolor{Gray}{gray}{0.85}
\definecolor{lightgray}{gray}{0.8}

\usepackage{caption}
\usepackage{subcaption}
\newcommand{\nm}[1]{\hline\multicolumn{1}{c}{\cellcolor{black} { {\bf 
\textcolor{white}{#1}}}}}

\usepackage[linesnumbered,ruled,vlined]{algorithm2e}
\SetKwProg{Fn}{Function}{}{}

\usepackage{amsmath}

\usepackage[framed]{ntheorem}
\usepackage{framed}
\usepackage{tikz}
\usetikzlibrary{shadows}
\theoremclass{Lesson}
\theoremstyle{break}

\tikzstyle{thmbox} = [rectangle, rounded corners, draw=black,
fill=Gray!20, drop shadow={fill=black, opacity=1}]

\newshadedtheorem{lesson}{Finding}

\newcommand{\quart}[4]{\begin{picture}(80,4)
{\color{black}\put(#3,2){\circle*{4}}\put(#1,2){\line(1,0){#2}}}\end{picture}}

\newcommand{\tion}[1]{\S\ref{tion:#1}}

\usepackage{color}
\usepackage{xspace}
\definecolor{ScarletRed}{rgb}{0.80,0.00,0.00}

\begin{document}

\title{Assessing Practitioner Beliefs about Software Engineering}

\author{N.C. Shrikanth \and
William Nichols \and \\Fahmid Morshed Fahid \and Tim Menzies 
}

\institute{N.C. Shrikanth, Fahmid Morshed Fahid, and Tim Menzies \at
Department of Computer Science, North Carolina State University, Raleigh, NC, USA. \\
\email{nc.shrikanth@gmail.com, timm@ieee.org, and ffahid@ncsu.edu}
\\ \\
William Nichols \at
Software Engineering Institute, Carnegie Mellon University, Pittsburgh, PA, USA. \\
\email{wrn@sei.cmu.edu}
}

\date{\textbf{Publication reference}: https://link.springer.com/article/10.1007/s10664-021-09957-5\\
\textbf{DOI}: https://doi.org/10.1007/s10664-021-09957-5\\
Accepted: 26 February 2021 }

\maketitle

\begin{abstract}
Software engineering is a highly dynamic discipline.
Hence, as times change, so too might our beliefs about
core processes in this field.

This paper checks some five beliefs that originated in the past decades that comment on the
relationships between (i)~developer productivity;
(ii)~software quality and (iii)~years of developer experience. 

Using data collected from 1,356 developers in the period 1995 to 2006, 
we found support for only one of the five beliefs titled ``\textit{Quality entails productivity}.'' We found no clear support for four other beliefs based on programming languages and software developers. However, from the sporadic evidence of the four other beliefs, we learned that a narrow scope could delude practitioners in misinterpreting certain effects to hold in their day-to-day work. Lastly, through an aggregated view of assessing the five beliefs, we find programming languages act as a confounding factor for developer productivity and software quality.  

Thus the overall message of this work is that it is both important and possible to revisit old beliefs in software engineering. Researchers and practitioners should routinely retest old beliefs. 

\end{abstract}

\newpage

\section{Introduction}\label{sect:introduction}

\begin{raggedleft}
``\textit{
Though deeply learned, unflecked by fault, 'tis rare\\ to see
when closely scanned, a man from all unwisdom free.}''\\ 
-- Valluvar's sacred couplet (translated, 1886, G.U. Pope ~\cite{pope1999sacred,thirukkural})

~\\

\end{raggedleft}

Ideally, practitioners and researchers in Software Engineering (SE) learn
lessons from the past
in order to better manage their
future projects.
But while many researchers record those beliefs~\cite{wan2018perceptions,xia2019practitioners,xia2017developers,zou2018practitioners}, 
very little is currently being done to verify the veracity
of those beliefs.

We assert that it is important to
quantitatively assess SE beliefs, as 
such beliefs are used by
\begin{itemize}
\item
{\em 
Practitioners} when they justify design or process decisions; e.g. ``better not use goto statements in our code'';
\item {\em Managers} to justify purchases or training programs or hiring decisions; e.g. ``test-driven development processes are best'';
\item and {\em Researchers} as they select what issues they should explore next; e.g. ``it is better to remove more bugs, earlier in the life-cycle,
since the longer they stay in the code, the more expensive it becomes to remove them.''
\end{itemize}
But the justification for such beliefs may be weak. 
Nagappan et al. recently
rechecked and rejected Dijkstra's famous comment that goto is necessarily considered harmful~\cite{nagappan2015empirical}.
As to early bug removal, 
Menzies et al. looked for evidence about whether or not ``the longer a bug remains in the system, the
exponentially more costly it becomes to fix.'' An extensive literature survey found only ten papers that actually experimented with this issue, of which five did, and five did not support this belief~\cite{menzies2017delayed}.
Further, Fucci et al. reviewed numerous studies on test-driven development and
found no evidence of an advantage from writing tests before writing code~\cite{Fucci17}. To say the least, this result is very different from numerous prior claims~\cite{beck03}, 

More generally,
Devanbu et al. reported at ICSE'16 just how widely practitioner beliefs at Microsoft diverged from each other and from the existing empirical evidence ~\cite{devanbu2016belief}. Also,
Shrikanth and Menzies reason that discrepancy between practitioners and empirical evidence by documenting the poverty of evidence for numerous defect prediction beliefs in dozens of software projects ~\cite{shrikanth2020assessing}.

Motivated by the above examples, in this paper, we:
\bi
\item
{\em Determine what large data sources exist}. Since 1995, our second author (Nichols) has been tutoring data collection methods for developers.
As part of that work, he has collected data from ten tasks
assigned to 1,356 developers. 
In all, we have data from
5,424 completed tasks.
\item {\em Check how that data
comments on known catalogs of SE beliefs.} For this paper, we used
the 2003 textbook \textit{A handbook of software and systems engineering: Empirical observations, laws, and theories}~\cite{endres2003handbook} by Albert Endres \& Dieter Rombach. That book documents dozens of
SE hypotheses, laws, and theories. 
\ei
For a variety of reasons,
this paper only explores the five Endres and Rombach beliefs listed in \tbl{beliefs}. Those reasons are:
\bi
\item No single article
could explore all the beliefs
recorded by Endres and Rombach.
\item
The data used in this study provided by Nichols (second author of this paper) et al. ~\cite{psp_data} could only comment on a subset of the Endres and Rombach beliefs. As to beliefs such as ``\textit{Prototyping (significantly) reduces requirement and design errors, especially
for user interfaces}'' (Boehm’s second Law) or ``\textit{Screen pointing-time is a function of distance and width}'' (Fitts–Shneiderman law) that would require a different data source to assess. 

\item
Of the remaining beliefs, we found that five were most widely-cited.
For example, one of the original SMALLTALK papers~\cite{goldberg1983smalltalk} (cited 7,430 times) motivates its work using the ``Dahl-Goldberg'' hypothesis listed in 
\tbl{beliefs}. Also the paper propose the ``Apprentice's Law''~\cite{Norman:1993:TMU:200550} has been cited 4,390 times. 
The remaining three beliefs are all referenced in the \textit{The Mythical Man-Month}~\cite{brooks1995mythical} and the 
famous 1987 article \textit{No silver bullet} ~\cite{brooks1987no}.
These two words are cited 8,649 and 5,085 times, respectively\footnote{All the citation counts in this bullet item were collected from Google Scholar, December, 2019.}. 
\ei

\begin{table}[!t]
\footnotesize
\centering
\caption{Beliefs studied in this paper.}
\begin{tabular}{|l|p{6cm}|p{5.7cm}|p{1cm}|}
\hline
\rowcolor[HTML]{EFEFEF} 
\textbf{\#} & \multicolumn{1}{c|}{\cellcolor[HTML]{EFEFEF}\textbf{Belief}} & \textbf{Conceived} \\ \hline
\textbf{1} & \textit{Productivity and reliability depend on the length of a program's text, independent of language level used.} & Corbat\'o's law ~\cite{corbato1969pl} (1969) \\ \hline
\textbf{2} & \textit{Object-oriented programming reduces errors and encourages reuse.} & Dahl-Goldberg Hypothesis ~\cite{dahl2001class, goldberg1983smalltalk} (1967 \& 1989) \\ \hline
\textbf{3} & \textit{Quality entails productivity.} & Mills-Jones Hypothesis ~\cite{mills1983software, cobb1990engineering} (1983 \& 1990) \\ \hline
\textbf{4} & \textit{Individual developer performance varies considerably.} & Sackman's Second law ~\cite{sackman1966exploratory} (1968) \\ \hline
\textbf{5} & \textit{It takes 5000 hours to turn a novice into an expert.} & Apprentice's law ~\cite{Norman:1993:TMU:200550} (1993) \\ \hline
\end{tabular}
\label{tbl:beliefs}
\end{table}
\begin{figure}[h]
\begin{center}
\includegraphics[width=2.5in,keepaspectratio]{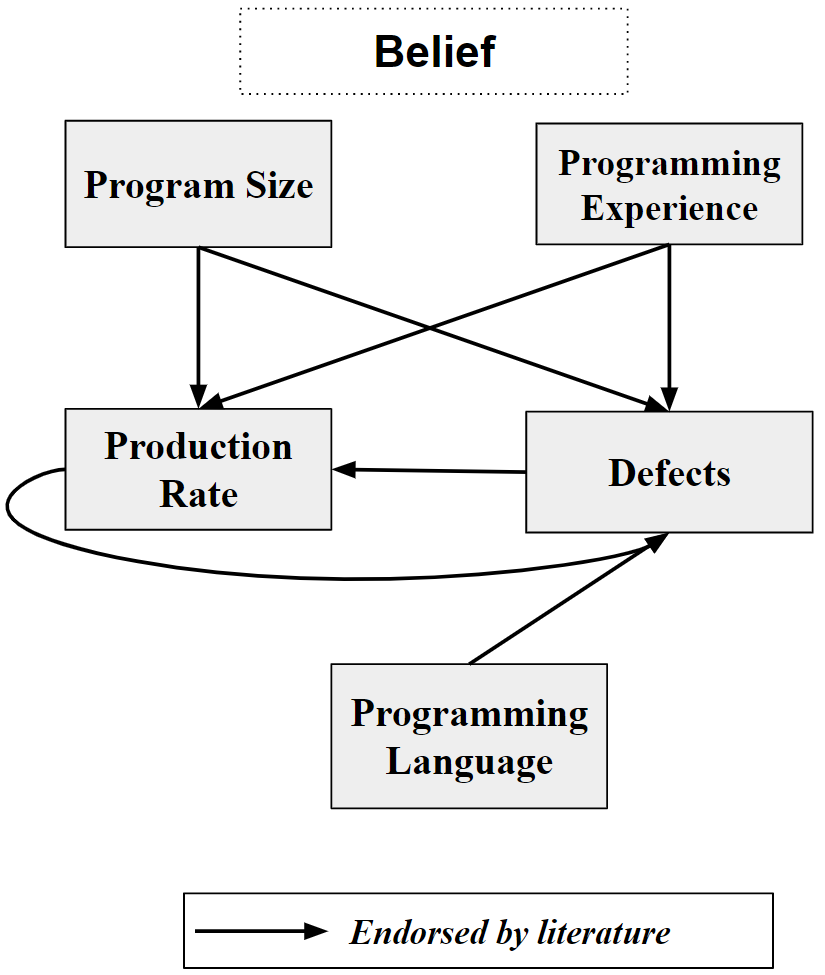}
\caption{A summary of beliefs in \tbl{beliefs} is shown here. Beliefs are broken into their entities and edges are drawn between two entities to acknowledge the presence of an effect as reported in SE literature. Strength of that effect using the data we collected is assessed in \tion{assessing} and this figure is updated (re-drawn) as per the current evidence later in \fig{triangulate}. }
\label{fig:triangulate_question}
\end{center}
\end{figure}
Finally, there is some coherence between the five beliefs we selected. Specifically, they explore aspects of the entities of \fig{triangulate_question}. That is to say, in theory, we could learn more from a summation of these beliefs than from just from a separate
study of each of them. 
Specifically, after studying the data about these five beliefs, we can ask and answer three research questions:

\begin{tcolorbox}
\textbf{RQ1: Why beliefs diverge among practitioners?}

\textit{Apart from belief 3 titled ``\textit{Quality entails productivity}'', none of the other beliefs are supported. } 
\end{tcolorbox}

Next, we ask:\\

\begin{tcolorbox}

\textbf{RQ2: What is the relationship between Productivity, Quality, and Expertise?}

\textit{A focus on quality, early in the project life-cycle minimizes rework but programming experience neither improves production rate nor mitigates defects.} 
\end{tcolorbox}

Finally, we ask:\\

\begin{tcolorbox}\textbf{RQ3: What impacts Productivity,  and Quality?}

Programming languages --- \textit{C\# and VB developers wrote programs with fewer defects. Specifically, C\# developers were most productive (among four other groups of developers who wrote programs in C, C++, Java, and VB).}

\end{tcolorbox}

The contributions of this paper are:
\bi
\item 
A replication study: We assess five SE beliefs to understand the widespread relevance of a large disconnect between SE beliefs and the actual evidence in practice.
\item
Prior publications such as ~\cite{devanbu2016belief, kochhar2016practitioners} report the disconnect between practitioners and empirical evidence, which is important, but only a few ~\cite{shrikanth2020assessing} extends to offer an explanation for that disconnect. We highlight such disconnects exist even among decades-old SE beliefs based on developer productivity, expertise, and software quality.

\item Importantly, our advice to practitioners is not to dwell into \textit{years of developer experience} but value some \textit{programming languages} over others. We also suggest practitioners focus on quality right from the early stages of a project, preferably adhering to a disciplined process. 

\item The data is publicly available~\cite{psp_data} and the results of this study are reproducible. The reproduction package is available  here~\footnote{\url{https://doi.org/10.5281/zenodo.4553435}}.

\ei

The rest of this paper is structured as follows. \tion{related} relates the current work to the prior literature as per the research questions we ask. Then in \tion{methodology} \& \tion{sem}, we discuss the choice of our datasets statistical tests, measures, and terminologies needed for assessment in \tion{assessing}, where we detail the modeling of the beliefs. Next, we discuss the results of our assessment in \tion{discussion}. \tion{threats} discusses the reliability of our findings. Lastly, we summarize in \tion{conclusion} and provide takeaways for practice in \tion{ifp}.

\section{Five Beliefs}\label{tion:related}

This section describes the beliefs
explored in this paper (and the next section describes the data we used to explore those beliefs).

\subsection{Quality:} Belief 2 claims that ``\textit{Object-oriented programming reduces errors and encourages reuse}''; i.e., some groups of programming languages induce more defects than others. In the literature, there is some support for this claim:
\bi
\item
Ray et al. analyzed Open Source (OS) projects and found a modest but significant effect of programming languages affecting software quality~\cite{ray2014large}.
\item
Kochhar et al. ~\cite{kochhar2016large, bissyande2013popularity} showed some languages used together (interoperability) with other languages induced defects.
\item
Bhattacharya and Neamtiu~\cite{bhattacharya2011assessing} argue that C++ is a better choice than C for both software quality and developer productivity.

\item Mondel et al. empirically assessed four beliefs related to systems testing. They found evidence for an old belief based on a more reused code to be harmful ~\cite{thomas1997analysis} in one of the two organizations they assessed ~\cite{monden2017examining}.

\ei

Belief 3 claims that ``\textit{Quality entails productivity}''; i.e., this belief implies a relationship between quality and productivity. Mills by applying Cleanroom Software Engineering, showed the possibility of simultaneous productivity and software quality improvements in both commercial and research projects ~\cite{mills1983software}.

\subsection{Productivity:} \label{tion:productivity}

Much prior work~\cite{sackman1966exploratory,kersten2006using,latoza2020explicit} in the past decades studied developer productivity. Belief 1 titled ``\textit{Productivity and reliability depend on the length of a program’s text, independent of language level used}'' implies that Lines of Code (LOC) is a better indicator of software quality and productivity than some programming languages. To the best of our knowledge, this 1969 belief is not well explored in the past. Compared to the late '60s, practitioners now write code in numerous programming languages using tools (like Integrated Development Environments) to catalyze software development. Thus it is essential to revisit the claimed effect. 

Interestingly, some researchers acknowledge the widely held belief that some good developers are much better (almost 10X) than many poor developers ~\cite{sackman1966exploratory}. Belief 4 is centered around the belief titled ``\textit{Individual developer performance varies considerably}.'' On related lines of thought, using the same data set, Nichols pointed out that a developer who is productive in one task is not necessarily productive in another ~\cite{nichols2019end}. That result warns us that even if we do find a \textit{hero}~\cite{agrawal2018we} developer, they may not remain \textit{heroes} consistently. Thus the focus should be to answer whether this productivity variance also impacts software quality? If it does not, then practitioners can confidently withdraw their large appeal around these moderate productivity variances in practice.

While exploring literature on developer productivity, we also note a common debate on universal productivity metric,
\bi
\item
In one study, Vasilescu et al. measured productivity as the number of pull requests to show productivity improvements through Continuous Integration practice in the GitHub arena ~\cite{vasilescu2015quality}.
\item
In another recent study, Murphy et al. showed non-technical factors (self-rated metric) were good predictors for productivity ~\cite{murphy2019predicts}.
\item
Suggestions about how to augment traditional measures such as incorporating rework time were also discussed in the past ~\cite{paulk2006factors}. 
\ei

Since all the above productivity measures discussed have their limitations, we lean towards the most prevalent measure, `production rate' (program size over time) used in the literature. The list of measures used in this study refers to \tion{sem}.

\subsection{Expertise:} Two common beliefs are experts perform the same task better (higher quality and meet deadlines) than novices and that expertise is built over time. The differences between experts and novices are discussed in various domains ~\cite{ericsson2004deliberate,ericsson1993role}. In SE back in 1985, Wiedenbeck considered 20 developers in two equal groups of 10 and found the expert group to be significantly better in certain programming sentence identification tasks than the other novice group. The expert group had 20,000 hours (mean) experience in their programming languages, whereas the novice population had as little as 500 hours (mean).  

Although some studies have highlighted there is more than just years of experience to expertise~\cite{baltes2018towards}, we think it is important to revisit prevalent beliefs. Especially belief 5 titled ``\textit{It takes 5000 hours to turn a novice into an expert}''; as it is known to influence software quality and developer productivity. For example, a 2014 TSE article by Bergersen et al. claimed that the first few years of experience correlated with developer performance. But later, a 2017 EMSE article by Dieste et al. found years of experience to be a poor predictor of developer productivity, and quality ~\cite{dieste2017empirical}.


Our work is similar to ~\cite{devanbu2016belief,shrikanth2020assessing} where we too assess various beliefs in an empirical study, but we differ from them in the following ways: 
\bi 
\item The truisms we assess have influenced numerous SE articles as discussed earlier in~\tion{related}. The beliefs we assess are not specific to a particular space like defect-prediction metrics as in ~\cite{shrikanth2020assessing} but also extend to other SE entities such as developer productivity and expertise.

\item We observe variations in entities of beliefs such as developer productivity, defects, and years of developer experience among different programming languages. The results of that observation can help managers to prefer some programming languages over another.

\item Although some Open Source Software systems (OSS) lessons may extend to practice; this work looks for evidence in tasks completed by developers from industries of various domains. The generalizability of our results is discussed in \tion{threats}. 

\ei

\section{Data}\label{tion:methodology}

In this section, we discuss the source and nature of the data while detailing the collection framework. Then we detail statistical tests and SE measures used to answer our RQ's. 

In summary, our data comes from a decades-long training program. The consultants from the Software Engineering Institute (SEI, based in Pittsburgh, USA)
traveled around the world to train developers in personal data collection. This ``Personal Software Process'' (or PSP) ~\cite{Humphrey1995}
is based on a belief that a disciplined process can improve productivity and quality ~\cite{paulk2010impact}. Specifically, if a practitioner uses PSP, they are encouraged to guess how long some tasks will take and then explain any differences between the predicted and actual effort. 

There are several reasons to use this data. Firstly, it is a minimal intrusion into the actual development work of practitioners. With the support of the right tools (e.g., with the tools from the SEI), practitioners spend less than 20 minutes per day on the PSP data collection. Hence, PSP can generate accurate and insightful records of actual developer activity~\cite{paulk2010impact,vallespir2016quality,paulk2006factors,paulk2005empirical,nichols2019end}.

\begin{table}[h]
\centering
\normalsize
\caption{Count of Engineers (Developers) by Domain}
\label{tab:PSP Classes}
\begin{tabular}{|l|c|l|c|}
\hline
\rowcolor[HTML]{EFEFEF} \textbf{Product Domain} & \textbf{Number} & \textbf{Product Domain} & \textbf{Number} \\ \hline
Software Services &    378     & Telecom &    92 \\ \hline
Business IT &    351     & Financial &    68 \\ \hline
Automation\&Control &    112     & Government &    66 \\ \hline
Accounting Software &    99     & Embedded &    55 \\ \hline
Consumer Electronics &    99     & Aerospace &    51 \\ \hline
Automotive &    97     & Other &    319 \\ \hline
\end{tabular}
\label{tbl:domain}
\end{table}

\begin{figure*}[!t]
\begin{center}

\begin{center}
\frame{\includegraphics[keepaspectratio,height=5.75cm]{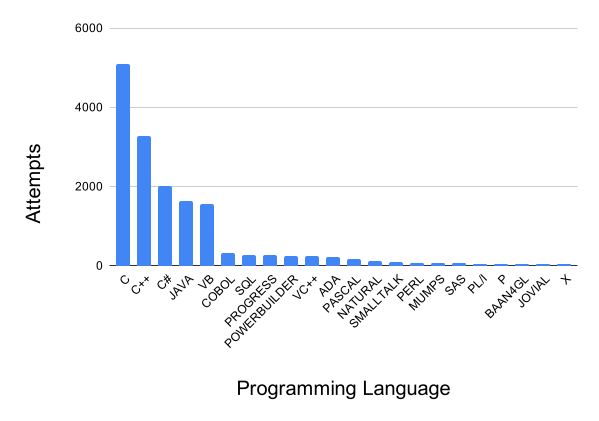}}
\end{center}

\caption{Distribution of all the tasks attempted \& completed by developers using a specific programming language. For a fair sample size comparison, we only consider tasks completed using C, C++, C\#, Java, and VB programming languages in this study. This bar-chart ignores 15 other languages that have been completed by less than 30 developers.}
\label{fig:pl_overview}
\end{center}
\end{figure*}

\begin{figure*}[!t]

\begin{center}
\frame{\includegraphics[keepaspectratio,height=7.25cm]{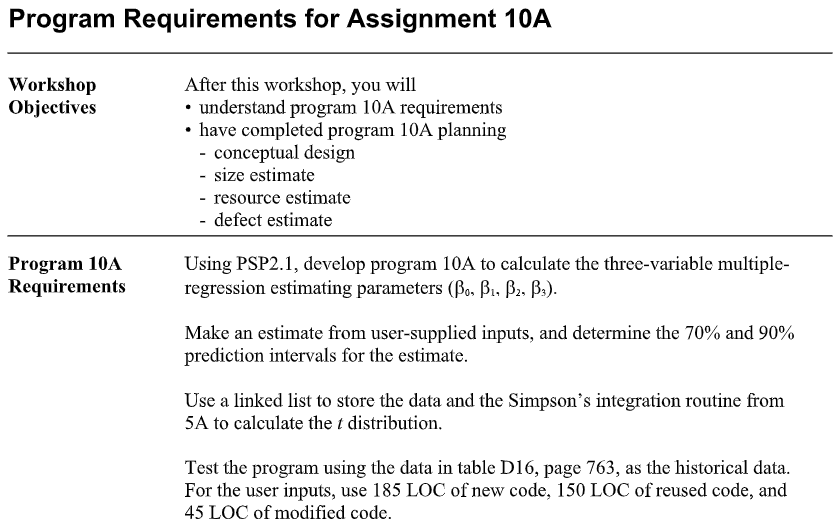}}
\end{center}
\caption{The task 10 (or assignment 10A) listed in \tbl{tasks} asks the developers to extend program 6 to calculate the three-parameter multiple-regression factors from a historical data set, then estimate the development effort with a 70\% to 90\% prediction intervals.
}
\label{fig:10A}
\end{figure*}

\begin{figure*}[h]
\centering

\begin{center}
\frame{\includegraphics[keepaspectratio,height=4.75cm]{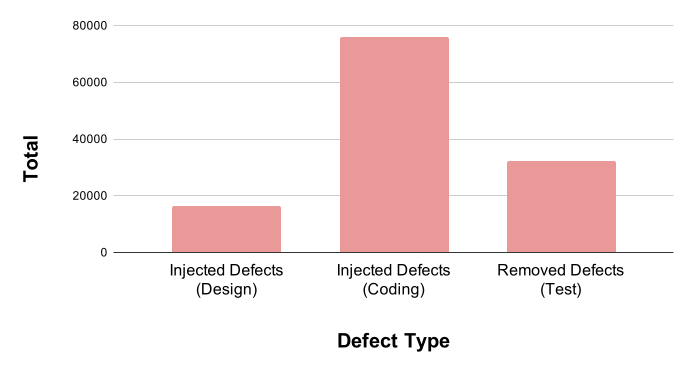}}
\end{center}

\caption{ Distribution of 124,521 defects recorded by developers while completing the 10 tasks using 5 programming languages C, C++, C\#, Java and VB.  Specifically 16,437 design and 75,927 coding defects were injected and 32,157 test defects were removed by 1,356 developers.}
\label{fig:where_defects}
\end{figure*}

Secondly, when SEI consultants train practitioners in PSP, they use a standard set of ten tasks.
The course is taught over 10 class days, with one week focused on measurement and estimation, and the second week focused on reviews, design, and quality (and there was typically a minimum two-week gap between weeks one and two). 
Hence, we have data on thousands of developers doing the same set of tasks, using a wide variety of programming methods and tools. For an overview of that data:
\bi
\item
\tbl{domain} lists the thousands of developers who have had this PSP training, along with the kind of software they usually develop.
\item
\fig{pl_overview} lists the languages used by attendees as they tried to complete the ten programming tasks. 
\item
\tbl{tasks} sorts the ten tasks (labeled from 1 to 10) from simplest (at level ``0'') to hardest (at level ``2''). Small dice of a 20-page task 10 specification is presented in \fig{10A}. Concise requirements for task 10 include writing programs to 
\bi
\item Read a table of historical data using the linked list from task 1
\item Write a multiple regression solver to estimate the regression parameters 
\item From user-supplied values of estimates for new LOC, reused LOC, and modified LOC compute the expected effort and prediction interval
\item Print out the results
\ei
\item 
\fig{where_defects} lists the tens of thousands of defects recorded during the PSP training tasks.
\ei

\begin{table}[!b]

\centering
\caption{Overview of the data set composed of 10 tasks at various levels completed in one of the five programming languages we considered. Level 2 (the rows are shown in gray) task have the highest complexity than its earlier levels.}
\normalsize
\begin{tabular}{|l|r|r|c|}
\hline
\rowcolor[HTML]{EFEFEF} 
{\color[HTML]{000000} \textbf{Level}} &
{\color[HTML]{000000} \textbf{Task}} &
{\color[HTML]{000000} \textbf{Developer Attempts}} &
{\color[HTML]{000000} \textbf{Programming Languages}} \\ \hline
& 1 & 1,356 & \\ \cline{2-3}
& 2 & 1,356 & \\ \cline{2-3}
\multirow{-3}{*}{0} & 3 & 1,356 & \\ \cline{1-3}
& 4 & 1,356 & \\ \cline{2-3}
& 5 & 1,356 & \\ \cline{2-3}
\multirow{-3}{*}{1} & 6 & 1,356 & \\ \cline{1-3} 
\cellcolor[HTML]{EFEFEF} & \cellcolor[HTML]{EFEFEF}7 & \cellcolor[HTML]{EFEFEF}1,356 & \\ \cline{2-3}
\cellcolor[HTML]{EFEFEF} & \cellcolor[HTML]{EFEFEF}8 & \cellcolor[HTML]{EFEFEF}1,356 & \\ \cline{2-3}
\cellcolor[HTML]{EFEFEF} & \cellcolor[HTML]{EFEFEF}9 & \cellcolor[HTML]{EFEFEF}1,356 & \\ \cline{2-3}
\multirow{-4}{*}{\cellcolor[HTML]{EFEFEF}2} &
\cellcolor[HTML]{EFEFEF}10 &
\cellcolor[HTML]{EFEFEF}1,356 &
\multirow{-10}{*}{C, C++, C\#, Java and VB} \\ \hline
\end{tabular}
\label{tbl:tasks}
\end{table}

Thirdly, this PSP data comes from industrial practitioners from the world. 
The PSP classes were taught in the US, Japan, Korea, Australia, Mexico, Sweden, Germany, the United Kingdom, the Netherlands, and India. Class size ranged from 1 to 20 developers, with a mean of 10.4 and an interquartile range of 7 to 14. Only about 3.2\% of subjects (123 of 3,832) were from a university setting, while most of the classes, 361 of 373, were taught in the industry to practicing software developers. Early adopters included Air Force, ABB, Honeywell, Allied Signal, Boeing, and Microsoft. 

Fourthly, this is high-fidelity data. A study of PSP data collection by Disney and Johnson \cite{Johnson1999} using 10 developers who wrote 89 programs found that manual collection and calculations on paper led to a 5\% error rate, mostly in derived calculations or transcription errors.
One explanation for this low error rate is the way the data was collected. The SEI authorized instructors to review each developer's (student)  PSP data as a required criterion for the successful completion of the task. Grading rubrics included self-consistency checks, checks to ensure that estimates and actuals are consistent with historical data, and comparisons with peers for data from each sub-process. Developers are also shown class summaries for comparison to their peers. Hence, various studies~\cite{rombach2008teaching, Vallespir2012, Grazioli2013} have found the data to be very accurate.

As to the nature of the ten programming tasks:
\bi
\item
They varied slightly in size, difficulty, and complexity. 
\item
They were chosen to be sufficiently difficult to generate useful data on estimation, effort, size, and defects and could typically be completed in an afternoon with 100 to 200 Lines of Code in a 3rd Generation language. 
\item Two programs were dedicated to counting program size; the remainder were primarily statistical, including regression, multiple regression, a Simpson's rule integral, Chi function, Student's T function, and prediction intervals.
\item
Developers were not expected to be domain experts and were provided a specification package that included descriptions of necessary formulas, algorithms, required test cases, and numeric examples suitable for a developer with no specific statistical expertise. 
\ei

%
The developers were instructed to bring their own devices to the class with the understanding that they should be familiar with the development environment and use the programming language with which they were most comfortable.  We made it clear that this was a process course, not a programming course, and that their results depended upon not introducing confounders.  Students were also instructed that the purpose of the exercises was to produce a measurable amount of code, effort, and defects; therefore, they should not use library procedures. However, the use of primitive language functions such as square root, logs, and trig functions was expected. 

The developers collected their personal data for effort, size, and defects using the PSP data framework, which measures direct time in minutes and program size in new and changed lines of code. Developers were instructed to build solutions with incremental cycles of design, code, and test, selecting their own increment size, typically a component or feature of 25 to 50 lines of code. However, some developers could produce working programs in a single cycle; most used 3 to 5 cycles, depending on their solution size and complexity. For effort accounting, each increment was initially designed and coded (creation), reviewed (appraisal), followed by the compile and test (failure). All-time required to achieve a clean compile was attributed to compile. All rework necessary to get the tests to pass was attributed to the test. The accounting highlighted rework so that rework could be minimized. 

The developers counted all defects that escaped a development phase. A defect was defined as any change needed to correct the program that was discovered after a phase was considered to be complete.  Detection of defects was primarily during a personal review, a compile, or test. For example, a coding defect would typically be discovered during code review or compile but might escape into testing. Defect data included the fixed time, the type, the phase origin, and the discovery phase.

To help highlight rework, the phase was defined as a logical step, where a time of activity was the primary phase rather than the literal activity performed.  For example, coding must be followed by compile, then test. A defect in the test did not trigger a new accounting cycle. Any changes resulting from code and re-compile were attributed to the test.  IDE and static analysis tools required additional instruction for consistent accounting.

 The use of static code analysis was uncommon and prohibited until after the compile, and then it must be considered part of the compile phase. Compile was complete when all discovered defects were resolved.   IDE real-time syntax checking presented another unique condition, which students chose to use IDE  was not recorded. Those using an IDE were instructed to disable real-time syntax checking to have the maximum number of defects available for finding in the review or compile. We later abandoned the guidance to disable IDE checking because the individual baselines were sufficient for the course objectives. Nonetheless, the instruction to disable IDE checks was in effect for courses from which this data was taken.
%
 
 For accounting purposes, PSP categorizes the activities as Creation (Design, Code),  Appraisal ( Design review, code review), and failure (compile, test). These were logical phases rather than strict activities (i.e., A logical sequence is to design, code, compile, and test a piece of code).
 
For small programs, a waterfall was practicable, but most chose to proceed through the phases using incremental development.  If incremental,  each increment proceeded through the phases proceeded without reentry. For example,  fixing a bug in the test required some coding and re-compile, but effort and defects were assigned to the test. This accounting choice made rework more visible and allowed additional auditing of the data quality. 

Defects were mostly injected during a creation phase, design, or code, with a few injected while fixing another defect.
Defects were discovered during the compilation, execution of test cases, or by a personal review. The course design was to baseline the defects levels in compile and test then demonstrate the ability of review to find at least 60\% of those defects before escaping into a failure phase.

\begin{figure*}[!t]
\centering
\begin{center}
\frame{\includegraphics[keepaspectratio,height=5.75cm]{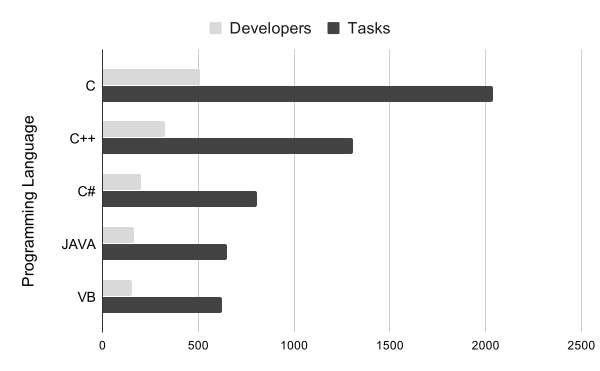}}
\end{center}
\caption{This bar-chart portrays the proportion of 1,356 developers completing 5,424 level 2 (labeled 7, 8, 9, and 10) tasks grouped by a specific programming language.}
\label{fig:pie_assign_lang}
\end{figure*}

\section{Methods} \label{tion:sem}

\subsection{Data Filtering}
We filtered the PSP data as follows:
\bi

\item
Although developers used numerous programming languages to complete the tasks, predominately 85\% of the developers used C, C++, C\#, Java, and Visual Basic (VB) programming languages as shown in \fig{pie_assign_lang}. We focus on these programming languages since they are very prevalent in the industry. 

\item
For simplicity in presentation belief, 1 and 4 use data only from level 2 tasks (labeled 7,8,9, and 10) listed in \tbl{tasks}. For all other beliefs (2, 3, and 5), we consider all the 10 tasks.

\item
Suiting to the nature of the beliefs, we use data from the appropriate type of defect in our analysis. The three types of defects we study are shown in \fig{where_defects}.
\ei

\subsection{Measures} 

We use the three SE measurements below to derive our conclusions while assessing the 5 beliefs we chose in this study.

\begin{tcolorbox}
\bi
\item[] \textbf{Program size (LOC)} = Lines of Code
\item[] \textbf{Production Rate} = LOC/Hour (\textit{Coding time})
\item[] \textbf{Quality (defects)} = Number of defects 
\item[] (\textit{Unless specified we consider defects injected in the coding phase. Other type of defects we analyze are defects injected in design phase and defects removed in testing phase.})

\ei
\end{tcolorbox}

As mentioned earlier, information per programming task such as the number of defects, program size, coding time, etc., is captured in practice by developers. To recollect, developers completed the 10 programming tasks of increasing complexity listed in \tbl{tasks}. They used various programming languages, as shown in \fig{pl_overview}, but largely using C, C++, C\#, Java, and VB, which we consider in this study. These traditional SE measures are used in these related studies ~\cite{wohlin2002prior,paulk2006factors}. 

\subsubsection{Threshold:} \label{tion:threshold}
 All the five beliefs we assess are constructed to compare different distributions of the measures like program size, production rate, and defects. 
One of us, Nichols, had previously applied structural equation modeling (SEM) on related problems to quantify effects but encountered modeling difficulties. The first issue was that SEM typically models linear effects, but the effort is the product of size and productivity.  This can be resolved by modeling productivity as the loading factor between size and effort, but the loading is not easily used as a separate variable.  Separately modeling productivity is possible, but since it is derived from other variables, those variables should not be included in the model because that induces spurious correlations.   A second issue was that the individual student seems to affect most other variables \cite{nichols2019end} with a non-normal distribution, including productivity. Neither the student effect nor other confounders are accounted for in the beliefs.   In summary, a unified parametric model seemed intractable, or at least premature.   
 We abandoned this approach as either unworkable or beyond our expertise.   Others, especially those with expertise involving interaction effects and moderation, are invited to use the provided data to apply SEM. Instead, we focused on alternate approaches. 

To cater to our experiment setup for beliefs 1, 2, 4, and 5 later in \tion{assessing} we employ the Scott-Knott test detailed in \tion{sk}. That test clusters and ranks two or more distributions. Belief 3 requires a method to check for linearity and to that we use Spearman's rank correlation (detailed in \tion{methodology}). From the usage of Spearman's $\rho$ (correlation co-efficient) in this SE literature ~\cite{zimmermann2007predicting} we derive the following ranges for $|\rho|$ :
\begin{tcolorbox}
\textbf{Correlation:}
\begin{itemize}
\item[$\ast$] 0.0 to 0.39 as \textit{no support}
\item[$\ast$] 0.4 to 0.49 as \textit{minimum/weak support}
\item[$\ast$] 0.5 to 0.59 as \textit{support}
\item[$\ast$] 0.6 to 0.69 as \textit{strong support}
\item[$\ast$] 0.7 to 1.00 as \textit{very strong support}\\
\item[]\textit{We acknowledge that these ranges are debatable. All the correlations we report are at the 99\% confidence level (ie., $p\_value < 0.01$).\\}
\end{itemize}
\textbf{Rank:}
\begin{itemize}
\item[$\ast$] \textit{Lower Scott-Knott rank indicates better production rate, better quality (fewer defects) and large program size (LOC) ie., Population distribution in Rank 1 is better than Rank 2.}
\item[$\ast$] \textit{Distributions placed in different ranks indicate significantly different population.}
\end{itemize}
\end{tcolorbox}

\subsection{Statistical Analysis}

To assess the beliefs and answer RQ1, RQ2, and RQ3, we need statistical tests that compute,

\begin{itemize}
\item \textit{Rank:} Clusters a list of populations to report significant differences.

\item \textit{Correlation:} Reports significant associations between two variables.
\end{itemize}

\subsubsection{Rank}\label{tion:sk}

Later in our experiments in \tion{assessing}, we compare populations of SE measures such as defects, production rate, and program size. Note populations may have the same median, but their distribution could be very different, hence to identify significant differences or rank among  many populations, we use the Scott\-Knott test recommended by Mittas et al. in TSE’13 ~\cite{Mittas13}. 

ScottKnott is a top-down bi-clustering approach used to rank different treatments; the treatment could be program size, production rate, defects, etc. This method sorts a list of $l$ treatments with $\mathit{ls}$ measurements by their median score. But before we sort a list of $l$ treatments, we normalize our data between [0, 1]. This is because the SE measures, like program size, defects, etc., do not typically fall between a fixed range to fit the quartile plots (later in \tion{assessing}). Thus to overcome this issue, we transform the list of $l$ treatments by applying min-max normalization, as shown below. Note this transformation does not impact the rank of the $l$ treatments in any way. 

\[ x'={\frac  {x-{\text{min}}(x)}{{\text{max}}(x)-{\text{min}}(x)}} \] 

Where,

\begin{itemize}
    \item $ {\text{max}}(x)$  is the global maximum ie., (largest value among the list of $l$ treatments)
     \item $ {\text{min}}(x)$  is the global minimum ie., (least value among the list of $l$ treatments)
\end{itemize}

The Scott-Knott approach then splits the normalized $l$ into sub-lists $m,n$ in order to maximize the expected value of differences in the observed performances before and after divisions. For lists $l,m,n$ of size $\mathit{ls},\mathit{ms},\mathit{ns}$ where $l=m\cup n$, the ``best'' division maximizes $E(\Delta)$; i.e.
the difference in the expected mean value
before and after the spit: 
\[E(\Delta)=\frac{ms}{ls}abs(m.\mu - l.\mu)^2 + \frac{ns}{ls}abs(n.\mu - l.\mu)^2\]

Notably, these techniques are preferred since they do not make Gaussian assumptions (non-parametric). To avoid ``small effects" with statistically significant results, we employ the conjunction of bootstrapping and A12 effect size test by Vargha and Delaney~\cite{vargha2000critique} for the hypothesis test H to check if \textit{m, n} are truly significantly different.

\subsubsection{Correlation} \label{tion:spear}

Spearman's rank correlation (a non-parametric test) assesses associations between two measures discussed earlier, for example, a correlation between production rate and software quality. We chose Spearman like some SE quality study ~\cite{d2010extensive} recommended to handle skewed data; further, it is unaffected by transformations (such as log, reciprocal,  square-root, etc.) on variables. 

The Spearman's rank correlation,
\mbox{$\rho = \frac{\text{cov}(X,Y)}{\sigma_x \sigma_y}$}
between two samples $X,Y$ 
(with means $\overline{x}$
and $\overline{y}$), as estimated using $x_i \in X$ and $y_i \in Y$ via
\[\rho = \frac{{}\sum_{i=1}^{n} (x_i - \overline{x})(y_i - \overline{y})}{\sqrt{\sum_{i=1}^{n} (x_i - \overline{x})^2(y_i - \overline{y})^2}}\]

We conclude using both the correlation coefficient ($\rho$) and its associated p\_value in all our experiments. The correlation coefficient ($\rho$) varies from +1, i.e., ranks are identical, to -1, i.e., ranks are the opposite, where 0 indicates no correlation.

\begin{itemize}
\item Higher $|\rho|$ value indicates strong evidence.
\item Lower $p\_value$ indicates the evidence is statistically significant.
\end{itemize}

\section{Assessing Beliefs} \label{tion:assessing}

In this section, for each belief listed in \tbl{beliefs}, we discuss the rationale, construct the experiment, and discuss the strength of the assessed belief. 

\begin{table*}[h]
\caption{ This table shows \textit{normalized} distributions of ``program size'', ``production rate'' and ``defects'' of level 2 task(s) ranked using Scott-Knott test (elucidated in \tion{sk}). Group 1 shows task 10 completed independently using C and C\# share similar LOC (the rows are shown in gray), whereas in the subsequent groups 2 and 3 we find the distributions of 10, C and 10, C\# to be significantly different.}
\centering
\frame{
{\footnotesize 
\begin{tabular}{l|l|l|r|r|l}

\nm{Program Size (LOC)}\\
\rowcolor[gray]{.9} \textbf{Group 1} & \textbf{Rank} & \textbf{Task, Language} & \textbf{Median} & \textbf{IQR} & \\
\hline 
\multirow{7}{3cm}{\centering  Level 2 tasks }

    &\cellcolor[gray]{.9}1 &   \cellcolor[gray]{.9}10, C\# &   \cellcolor[gray]{.9} 18 & \cellcolor[gray]{.9} 11 & \cellcolor[gray]{.9}\quart{13}{11}{18}{100} \\
    &\cellcolor[gray]{.9}1 &  \cellcolor[gray]{.9}10, C &  \cellcolor[gray]{.9}  17 & \cellcolor[gray]{.9} 10 & \cellcolor[gray]{.9} \quart{12}{10}{17}{100} \\
    & 2 &      9, C\# &    13 &  9 & \quart{9}{9}{13}{100} \\
    & 3 &      9, C &    11 &  7 & \quart{8}{7}{11}{100} \\
    & 4 &      7, C\# &    7 &  6 & \quart{5}{6}{7}{100} \\
    & 5 &      8, C\# &    6 &  5 & \quart{3}{5}{6}{100} \\
    & 5 &      7, C &    6 &  4 & \quart{4}{4}{6}{100} \\
    & 5 &      8, C &    6 &  5 & \quart{3}{5}{6}{100} \\
    
\nm{Production rate (LOC/hour)}\\ \hline
\rowcolor[gray]{.9} \textbf{Group 2} & \textbf{Rank} & \textbf{Task, Language} & \textbf{Median} & \textbf{IQR} & \\
\hline 
\multirow{3}{3cm}{\centering Task 10 completed using C and C\# }
& & & & & \\
&   1 &      10, C\# &    13 &  10 & \quart{9}{10}{13}{100} \\
& 2 &      10, C &    8 &  7 & \quart{5}{7}{8}{100} \\
& & & & & \\
\nm{Defects} \\ \hline
\rowcolor[gray]{.9}\textbf{Group 3} & \textbf{Rank} & \textbf{Task, Language} & \textbf{Median} & \textbf{IQR} & \\
\hline 
\multirow{3}{3cm}{\centering Task 10 completed using C and C\# }
& & & & & \\
 & 1 &      10, C\# &    5 &  7 & \quart{3}{7}{5}{100} \\
   &  2 &      10, C &    9 &  9 & \quart{5}{9}{9}{100} \\
& & & & & \\
\end{tabular}}}
\label{tbl:b1}
\end{table*} 

\subsection{Belief 1: Corbat\'o's law} \label{tion:b1}

This section discusses an effect reported in a 1969 paper by Corbat\'o ~\cite{corbato1969pl} that

\begin{center} \em Productivity and defects depend on the length of a program's text, independent of the language level used.
\end{center}

That is to say, (a)~longer programs tend to get more defects;
(b)~and this effect is not mitigated by newer generation languages.
Note that, if true, Corbat\'o's rule warns us that, by merely switching to a newer language:
\bi
\item
Defects cannot be reduced 
\item
And developers cannot be made more productive
\ei

%
To check this rule, we construct the experiment as follows:
\begin{itemize}
    \item[] (a) Group similar tasks that are written in both non-oo and oo programming languages.
    \item[] (b) In that group, select the same tasks that share similar LOC~\footnote{Following the belief statement, we use LOC (length of the program text) and not function points that share identical distribution. We compute production-rate (productivity) using LOC not just as defined in this book~\cite{endres2003handbook} (source of all the beliefs in this study) but also in this prominent studies~\cite {nguyen2011analysis,devanbu1996analytical}.}. 
    \item[] (c) Investigate production rate and defect distribution in those two groups.
\end{itemize}

Note, for this belief to be widely accepted, this rule should hold in  ``every'' oo-vs-non-oo programming language pair (such as C, C\# or C, C++, or C, Java or C, VB) that satisfy the above two experiment constraints (a and b). On the other hand, this belief cannot be endorsed if it does not show support even in any one of the oo-vs-non-oo language pairs.

To find the two groups that satisfy the experiment constraints, we started with the most complex level 2 tasks. If we do not find distributions that meet criterion (b), we would have moved to earlier level lesser complex tasks; those are in level 0 and 1. Lastly, in that group, we chose C and C\#  because Hejlsberg and Li et al. ~\cite{hejlsberg2006c,li2017empirical} assert that 
C and C\# are two programming languages at different ``levels,'' but more importantly, it satisfied the experiment constraints (a and b).  
%

High-level features help developers to write less code. For example, automatic memory management (garbage collection) is one of the numerous high-level features available in C\#. Automatic memory management can help developers to focus more on the assigned task's functional requirements rather than writing additional code to manage memory.

{\em Prediction:} If Corbat\'o was wrong, then we should see either

\bi
\item Production rates differ by programming language and/or
\item Defects differ by programming language.

\ei

\subsubsection{Result} 


\tbl{b1} shows our results in three groups program size, production rate, and defects. From this table, we make several observations.

\bi
\item

Program size distributions in group 1 reveal that tasks 8 and 10 completed using C and C\# share similar (same rank) LOC distribution.

\item 
Subsequently, in groups 2 and 3 (``production rate'' and ``defects''), we only focus on tasks 10, C, and 10, C\# results (the rows are shown in gray). We do that because (a)~we can remove the conflating factor of different LOCs (tasks 7 \& 9), and (b)~task 10 has higher LOC ranges than task 8, making it naturally a better choice for to carry further analysis.

\item
The focus of groups 2 and 3 (``production rate'' and ``defects'') on task 10 (chosen in the previous step) reveal developers who completed the task using C\# were more productive and induced fewer defects than those completed using C. 
\item Thus, as per Corbat\'o's Law, if only LOC matters and language level does not then task $10$ irrespective of whichever language (C or C\#) used should also portray similar production rate and defects distribution. However,
in \tbl{b1}, we observe a significant difference in the production rate and defects of these groups.
Thus we cannot ignore the level of a language as it impacts both developer production rate and defects.
\item 
Lastly, as mentioned earlier in \tion{b1}, given that this belief weakened with a C and C\# group, there is no need to assess this belief on remaining oo-vs-non-oo language pairs.

\ei

Accordingly, we say:
\begin{tcolorbox}{Belief 1: \textit{Our results contradicted Corbat\'o's law as with similar program size (LOC), tasks completed using C\# is significantly ``better'' (higher production rate and fewer defects) than using C.}}
\end{tcolorbox}

\subsection{Belief 2: Dahl-Goldberg hypothesis}\label{tion:b2}

This section discusses an effect reported in a 1983 paper by Dahl and Goldberg ~\cite{dahl2001class,goldberg1983smalltalk} that.

\begin{center}{\em Programs written using non-OO languages naturally induce more defects.}
\end{center}

If true, then programs written in OO languages like  Java should get fewer defects than written in C (non-OO).

To check this effect, we studied tasks completed by developers in 5 programming languages. Among those five languages VB, C\#, and Java support OO, whereas C does not support OO.   C++, often termed as an extension to C, does support OO; however, programmers may still write C like coding in C++. Hence, we do not premise our conclusion considering only C++ in our assessment. 

The rationale behind this belief, as discussed by Endres \& Rombach is that OO basically restricts the developer's freedom to prevent them from introducing unwanted defects. For example, information hiding (encapsulation), a concept in OO, is performed by developers while writing code to pacify software complexity and improve robustness. Hands-on, developers make use of access-modifiers such as \textit{private, protected} (in Java) to encapsulate certain complex parts of code. Further, modern OO languages such as C\# and Java do not easily expose low-level control or memory management for developers to manipulate them, but those features are readily available in C. 

To check if OO affect designs and the prevalence of  defects, we consider two types of defects from all the 10 tasks to assess this belief, they are: 
 
 \bi
 \item ``defects injected in design'' (design defects) and
 \item  ``defects injected in code'' (coding defects).
 
 \ei
 
Note this belief is not about examining the OO design paradigm, rather certain OO language features. As discussed by Endres and Rombach, OO languages offer certain features (such as automatic memory management, in-built libraries, etc.) that may prevent developers from injecting unwanted defects. In other words, one may still write a non-OO code using a OO language but take advantage of in-built features that OO languages offer.

{\em Prediction:} If Dahl \& Goldberg were wrong, then programming similar tasks using OO languages such as C\#, Java, and VB programs should have more or about the same range of defects compared to C.

\subsubsection{Result:} 

\tbl{b2} presents the ``defects (Code + Design)'' in two groups (programming languages and task 10). From this table, we make several observations.

\bi

\item The defect distributions in group 1 of developers using C\#, and VB (the rows are shown in gray) have fewer defects compared to those completed using Java, C, and C++.

\item Notably, tasks completed using Java that support OO show more defects similar to those written in C.

\item A focused analysis of defects in group 2 shows, task 10 completed in C\# and VB also share the least range of defects.

\item Defects are lower only in two of four languages that have some support for OO (C\# and VB), whereas Java and C++ (that support OO) portray significantly more defects similar to those written in C. Thus, we cannot endorse the Dahl-Goldberg hypothesis. 
\ei

\begin{table*}[h]
\caption{ This table shows \textit{normalized} distribution of ``defects (Coding + Design)'' in two different groups (programming languages and task 10). Using the Scott-Knott test (elucidated in \tion{sk}) we find tasks completed using C\# and VB in group 1 share least range of defects (the rows are shown in gray) compared to C, Java, and C++. The subsequent group 2 shows that pattern of similar low defects among C\# and VB for the most advanced task 10.}
\centering
\frame{
\begin{tabular}{l|l|l|r|r|l}
\nm{Defects (Coding + Design)}\\ \hline
\rowcolor[gray]{.9} \textbf{Group 1} & \textbf{Rank} & \textbf{Language} & \textbf{Median} & \textbf{IQR} & \\
\hline 
\multirow{5}{3cm}{\centering Programming Language}

  &  \cellcolor[gray]{.9}1 &\cellcolor[gray]{.9}VB &\cellcolor[gray]{.9}    5 &\cellcolor[gray]{.9}  5 &\cellcolor[gray]{.9} \quart{2}{5}{5}{100} \\
  &  \cellcolor[gray]{.9}1 &\cellcolor[gray]{.9}C\# &\cellcolor[gray]{.9}    5 &\cellcolor[gray]{.9}  5 &\cellcolor[gray]{.9} \quart{4}{5}{5}{100} \\
  &  2 &      C &    8 &  7 & \quart{5}{7}{8}{100} \\
  &   2 &      Java &    8 &  8 & \quart{5}{8}{8}{100} \\
  &   3 &      C++ &    11 &  10 & \quart{7}{10}{11}{100} \\ \hline
    
\rowcolor[gray]{.9} \textbf{Group 2} & \textbf{Rank} & \textbf{Task, Language} & \textbf{Median} & \textbf{IQR} & \\
\hline 
\multirow{5}{3cm}{\centering   Task 10 }

    & \cellcolor[gray]{.9}1 &\cellcolor[gray]{.9}10, C\#  &\cellcolor[gray]{.9}    5 &\cellcolor[gray]{.9}  5 &\cellcolor[gray]{.9} \quart{3}{5}{5}{100} \\
    & \cellcolor[gray]{.9}1 &\cellcolor[gray]{.9}10, VB &\cellcolor[gray]{.9}    6 &\cellcolor[gray]{.9}  6 &\cellcolor[gray]{.9} \quart{3}{6}{6}{100} \\
   &  2 &      10, Java &    8 &  11 & \quart{5}{11}{8}{100} \\
   &  2 &      10, C &    8 &  8 & \quart{5}{8}{8}{100} \\
   &  3 &      10, C++ &    11 &  13 & \quart{6}{13}{11}{100} \\
    
\end{tabular}
}

\label{tbl:b2}
\end{table*}

Accordingly, we say:
\begin{tcolorbox}
Belief 2: \textit{Programs written in OO are not necessarily less defect prone}
\end{tcolorbox}

\subsection{Belief 3: Mills-Jones hypothesis} \label{tion:b3}

This section discusses an effect from two papers by Mill \& Jones ~\cite{mills1983software, cobb1990engineering} in 1983 and 1990:

\begin{center}
{\em Quality entails productivity.} 
\end{center}

That is to say, a lack of early emphasis on quality in the project life-cycle will lead to a lot of rework (unproductive) and defective software. Mills showed that highly reliable software could be produced through cleanroom software engineering, which employs statistical-based independent testing ~\cite{mills1983software,mills1993cleanroom}. Disciplined processes such as cleanroom software engineering focus on quality right from the early stages of the project. Having such a focus minimizes unnecessary effort in the later stages of the project, unnecessary efforts like fixing defects in the final testing phase, which were undetected early in the project life cycle (such as coding or unit test).

To study this effect, we check for a linear trend between  ``code and design''   (defects injected during coding and design phases)  and ``test defects'' (defects escaped to testing phase) using the correlation test elucidated in \tion{sem}. We consider all 10 tasks (labeled 1 to 10) in \tbl{tasks} to gain more data points for the independent and dependant variables. Lastly, we export the significant correlation scores visually into a box-plot and discuss the strength of the observed trend based on the median. To achieve that, we do the following:

\bi
\item We capture the number of ``code and design defects,'' and the number of ``test defects'' for each task completed using a specific programming language.

\item Then, we correlate between the captured list of ``code and design defects'' and ``test defects'' across all the 10 tasks and export the correlation coefficient ($\rho$) values.

\item The above step results in 50 $\rho$ scores (10 tasks x 5 programming languages). We plot the exported scores (distributions) in \fig{b3}.
\ei

{\em Prediction:} If Mills \& Jones were wrong, then it could mean that managers need not invest in quality assurance activities early in their project life cycle.

\subsubsection{Result:}

\begin{figure}[!t]

\begin{center}
\frame{\includegraphics[keepaspectratio,height=4.75cm]{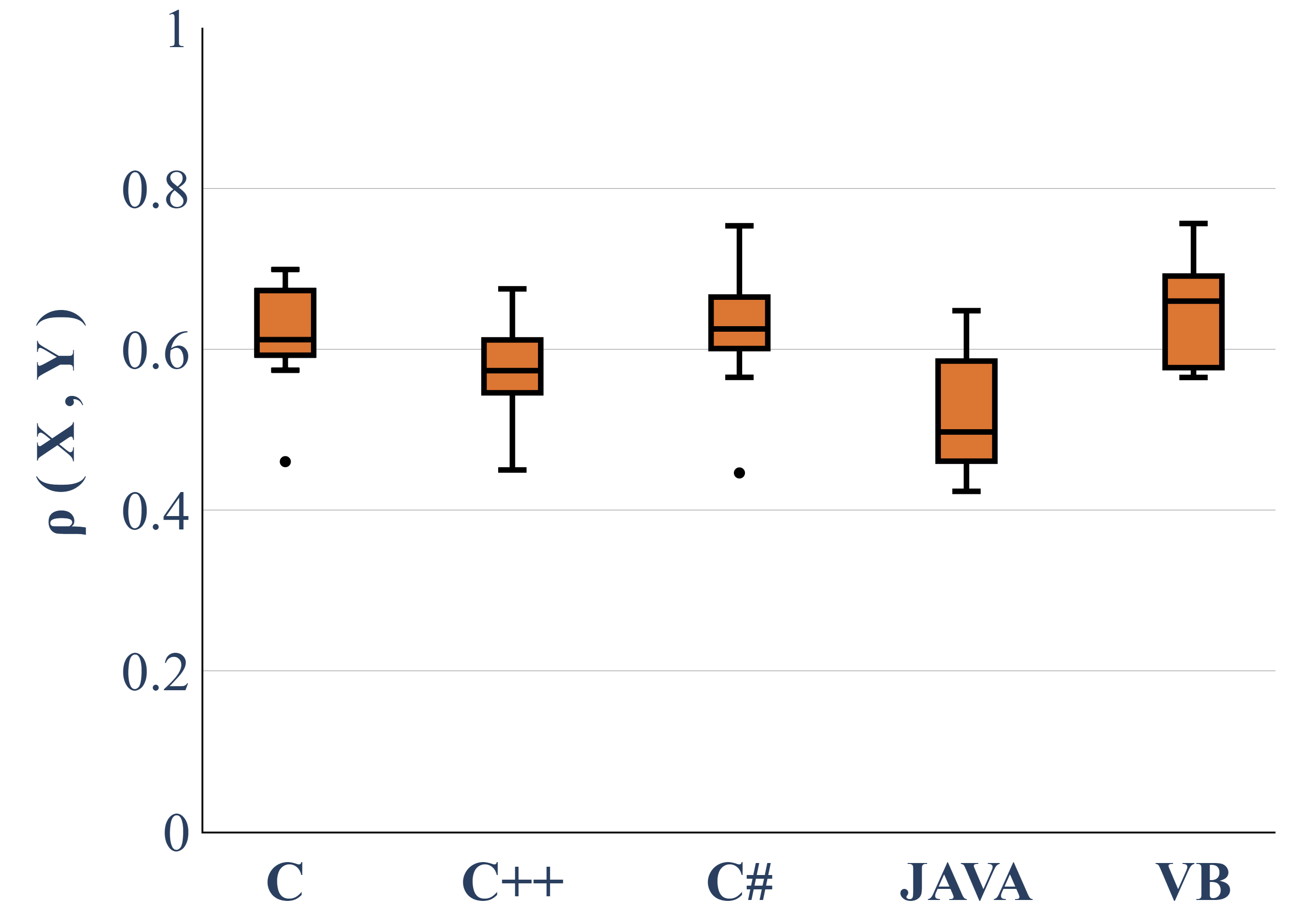}}
\end{center}
\caption{The box-plots in this chart show the distribution of correlation scores grouped by different programming language. The correlation is computed between      ``code and design defects''  (X) and ``test defects'' (Y) (computed as described in \tion{spear}) for each task (listed in \tbl{tasks}) completed using a specific programming language. 
}
\label{fig:b3}
\end{figure}
\fig{b3} presents a box-plot of all the exported correlation ($\rho$) scores grouped by programming language. We used all tasks (labeled 1 to 10) from \tbl{tasks}. From this figure, we make the following observations.

\bi

\item We find a median of  +0.5 ($\rho$) between ``code and design defects'' and ``test defects'' in Java, and in the remaining four programming languages we analyzed, the correlation is above   +0.6 ($\rho$).

\item An overall median correlation of    +0.6 ($\rho$)  considering all the five programming languages confirm rework will increase (more test defects) if there is a lack of emphasis on quality in the early stages. 


\ei

Accordingly, we say:
\begin{tcolorbox}{Belief 3: \textit{Emphasis on early quality  does minimize rework.}}
\end{tcolorbox}
That said, the strength of this support is not very strong $\rho$ (+0.7). 

\subsection{Belief 4: Sackman's second Law}

This section discusses an effect reported in a 1966 paper by Sackman et al. ~\cite{sackman1966exploratory}:

\begin{center}
{\em Individual developer performance varies considerably.}
\end{center}

That is to say, developer X is considerably ``better'' in completing a task than developer Y. By ``better'' we mean developer X writes more lines of code in less time than developer Y, and developer X's deliverable gets fewer defects than developer Y's deliverable.

Also, note that, if true, Sackman’s second Law warns us that:

\bi
\item Only some developers are productive and write quality code.
\ei

A variation between developers was rather a surprising finding by Sackman in 1966 as the objective of the original study was to compare productivity between online programming and offline programming ~\cite{sackman1966exploratory}. Endres \& Rombach also note that this effect is not extensively studied in the past few decades. They also offer some doubts concerning the small sample size and the statistical approach used in the original (Sackman's) study. Sackman's study considered only 12 developers, and their conclusion is based on extremes and not on the entire distribution. Note in this work; we compare large distributions of production rate and defect scores captured from over 1000 developers.

 Naturally, managers would prefer few high performers over many low performers, but recently (2019) Nichols using the same data, showed that a developer X who is productive in one task is not necessarily productive in another ~\cite{nichols2019end}. Thus we address the quality aspect of this belief.  To check whether large production rate variance among developers associate with more defects (low quality), we construct the experiment as follows:

\bi
\item We capture the number of ``code defects'' and the ``production rate'' for each task completed using a specific programming language. 

\item Then, we correlate between the captured list of ``code defects'' and ``production rate'' across all the ten tasks and export the correlation coefficient ($\rho$) values.  To expose any effect of the programming language in this analysis, we group the distributions by programming language, similar to the analysis earlier in \tion{b3} and later in \tion{b5}.

\begin{figure}[!t]

\begin{center}
\frame{\includegraphics[keepaspectratio,height=4.25cm]{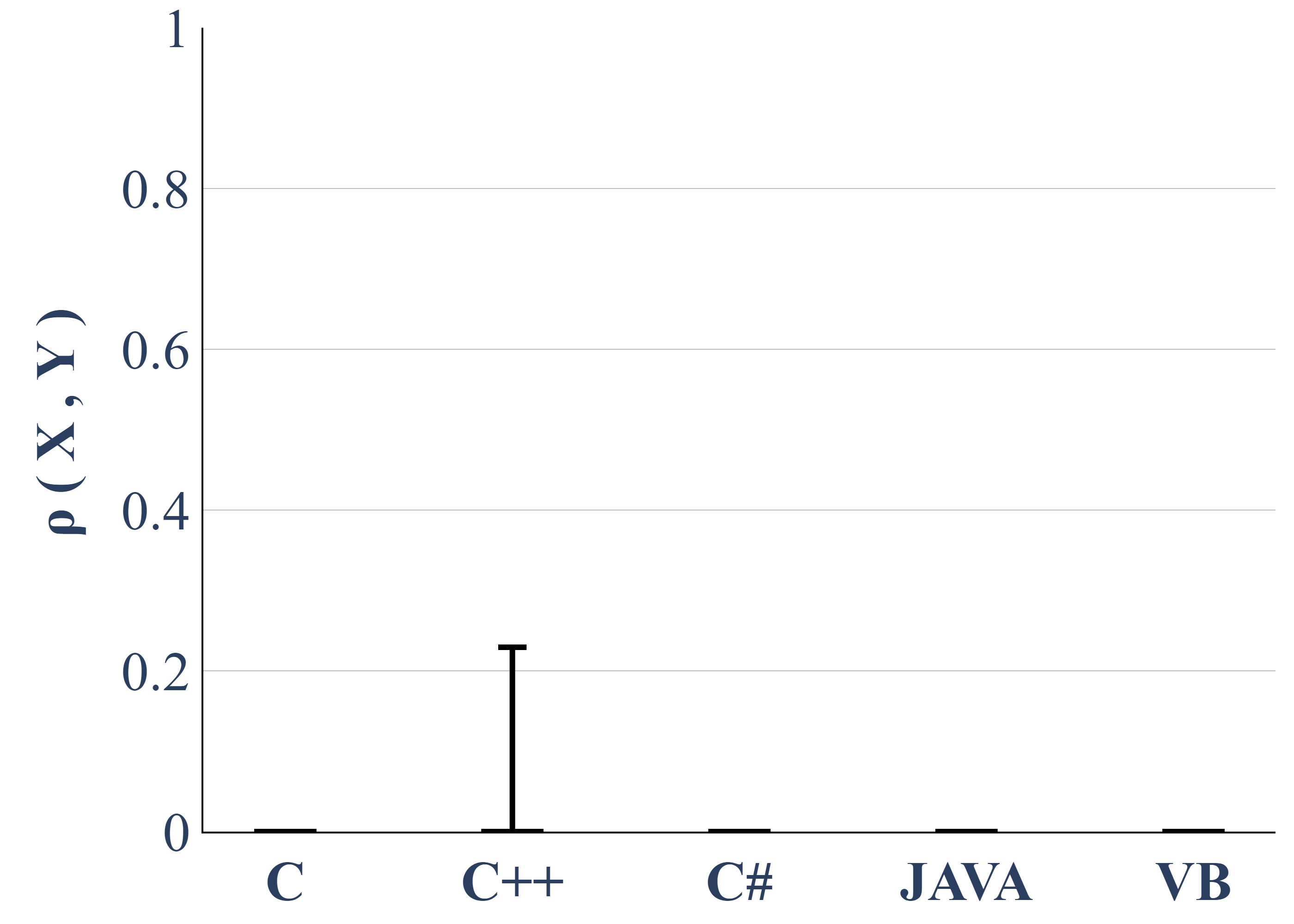}}
\end{center}
\caption{The box-plots in this chart show the distribution of correlation scores grouped by different programming language. The correlation is computed between   ``code defects'' (X)  and ``production rate'' (Y) (computed as described in \tion{spear}) for each task (listed in \tbl{tasks}) completed using a specific programming language. }
\label{fig:b4}
\end{figure}

\item The above step results in 50 $\rho$ scores (10 tasks x 5 programming languages); we plot the  exported scores (distributions) in \fig{b4}. 
\ei

{\em Prediction:} If Sackman was wrong, then practitioners may ease their large appeal towards some high-performing developers.

\subsubsection{Result:}

\fig{b4} presents a box-plot of all the exported correlation ($\rho$) scores grouped by programming language.  That figure reports a `0' ($\rho$) correlation score considering all the tasks (labeled 1 to 10 from \tbl{tasks}) and all the five programming languages independently. This result confirms the absence of a linear association between `production rate' and `code defects.'

Accordingly, we say:
\begin{tcolorbox}{Belief 4: \textit{Software Quality is not impacted by the variance in developer production rate.}}
\end{tcolorbox}

\subsection{Belief 5: Apprentice's Law} \label{tion:b5}

This section discusses an effect reported in a 1993 paper by Norman et al. ~\cite{Norman:1993:TMU:200550}; specifically:

\begin{center}
{\em It takes 5000 hours to turn a novice into an expert.}
\end{center}

To assess the effect of prolonged programming experience we analyze ``production rate'' and ``defects'' among the \textit{expert} and \textit{novice} groups. An expert is someone who is both knowledgeable and skilled in their field of work. An expert in this study is a developer who can complete the task on time (productive) with no defects (quality). 

Adopting from ~\cite{endres2003handbook} we map the 5000-hour threshold as follows:

\begin{itemize}
\item[] \textbf{expert} : $ \ge 3$ years of experience (or $ \ge 5000$ hours of programming experience)
\item[] \textbf{novice} : $ < 3$ years of experience (or $ < 5000$ hours of programming experience)
\end{itemize}

That is to say, (a) \textit{expert} developers induce \textit{less} defects than novices; (b) \textit{expert} developers are more productive in completing tasks than \textit{novice} developers. Note that, if true, the Apprentice’s Law warns us that, we should mistrust \textit{novices}
due to their lower  quality code. To check this,
we will analyze the distributions of ``production rate'' and ``defects'' among \textit{experts} and \textit{novices}.

All the specific tasks labeled 1 to 10 in \tbl{tasks} are new to the developers.  Nonetheless, while on real-world projects, developers will not get precisely similar task assignments, they may use similar skills including as follows:  applying the same language features; use iterations, conditionals, and subroutine interfaces;  use of data types and manipulation of data structures; developing test cases and debugging;  and so forth.   We test whether a developer, `X,' with four years of prior Java development experience, is better in completing the task, `T,' than another Java developer, `Y,' with less than a year of experience. Both developers `X' and `Y' are new to task `T' but have differing levels of experience using the underlying skills.

\subsubsection{Result:} 

\begin{figure*}[!t]
\centering
\begin{center}
\frame{\includegraphics[keepaspectratio,height=5.75cm]{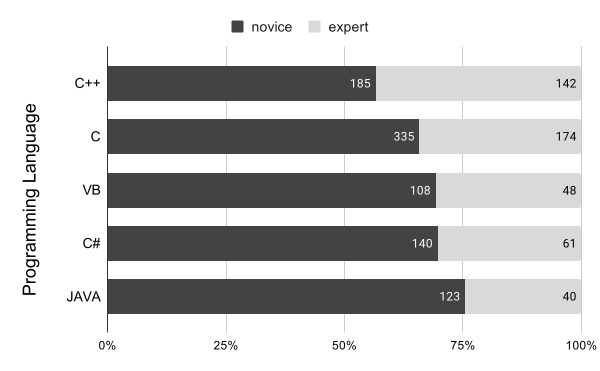}}
\end{center}
\caption{Proportion of \textit{expert} and \textit{novice} developers who completed all the 10 tasks listed in \tbl{tasks} using a specific programming language. 
} 
\label{fig:b5_bar_chart}
\end{figure*}

The ratio of \textit{expert} to \textit{novice} developers in our data is shown in \fig{b5_bar_chart}.

\begin{table*}[!t]
\caption{This table shows ``production rate'' and ``defects'' distributions of \textit{expert} and \textit{novice} developers in four groups. Using Scott-Knott test (elucidated in \tion{sk}) we find groups 1 and 2 indicate no difference in either production rate nor defects among \textit{expert} and \textit{novice} developers. On the other hand groups 3 and 4 portray significantly different distributions (as seen from different Scott-Knott ranks) among developers using different programming languages.}
\centering
\frame{
{\small 
\begin{tabular}{l|l|l|r|r|c}
\nm{Production rate (LOC/hour)}\\
\rowcolor[gray]{.9} \textbf{Group 1} & \textbf{Rank} & \textbf{Experience} & \textbf{Median} & \textbf{IQR} & \\
\hline 
\multirow{2}{3cm}{\centering All tasks}

    & 1 &      expert &    6 &  4 & \quart{4}{4}{6}{100} \\
    & 1 &      novice &    5 &  5 & \quart{3}{5}{5}{100} \\

\nm{Defects}\\
\rowcolor[gray]{.9} \textbf{Group 2} & \textbf{Rank} & \textbf{Experience} & \textbf{Median} & \textbf{IQR} & \\
\hline 
\multirow{2}{3cm}{\centering All tasks}
    & 1 &      expert &    6 &  6 & \quart{3}{6}{6}{100} \\
 & 1 &      novice &    5 &  6 & \quart{2}{6}{5}{100} \\

\nm{Production rate (LOC/hour)}\\
\rowcolor[gray]{.9} \textbf{Group 3} & \textbf{Rank} & \textbf{Language, Experience} & \textbf{Median} & \textbf{IQR} & \\
\hline 
\multirow{10}{3cm}{\centering Programming Language expertise}

    &1 &      C\#, novice &    8 &  6 & \quart{5}{6}{8}{100} \\
    &1 &      C\#, expert &    7 &  6 & \quart{5}{6}{7}{100} \\
    &2 &      C++, expert &    6 &  5 & \quart{4}{5}{6}{100} \\
    &2 &      Java, expert &    6 &  5 & \quart{4}{5}{6}{100} \\
    &2 &      VB, novice &    6 &  5 & \quart{4}{5}{6}{100} \\
    & 3 &      Java, novice &    5 &  4 & \quart{4}{4}{5}{100} \\
    &3 &      C++, novice &    5 &  5 & \quart{3}{5}{5}{100} \\
    &3 &      C, expert &    5 &  3 & \quart{3}{3}{5}{100} \\
    &3 &      VB, expert &    5 &  4 & \quart{3}{4}{5}{100} \\
    &3 &      C, novice &    5 &  4 & \quart{3}{4}{5}{100} \\ 
    
\nm{Defects} \\
\rowcolor[gray]{.9} \textbf{Group 4} & \textbf{Rank} & \textbf{Language, Experience} & \textbf{Median} & \textbf{IQR} & \\
\hline 
\multirow{10}{3cm}{\centering Programming Language expertise}

    & 1 &      C\#, novice &    3 &  3 & \quart{2}{3}{3}{100} \\
    & 1 &      VB, novice &    3 &  3 & \quart{2}{3}{3}{100} \\
    & 2 &      C\#, expert &    5 &  6 & \quart{2}{6}{5}{100} \\
    & 2 &      C, novice &    5 &  5 & \quart{3}{5}{5}{100} \\
    & 2 &      VB, expert &    5 &  5 & \quart{2}{5}{5}{100} \\
    & 3 &      Java, expert &    6 &  6 & \quart{3}{6}{6}{100} \\
    & 3 &      C, expert &    6 &  6 & \quart{3}{6}{6}{100} \\
    & 3 &      C++, novice &    6 &  7 & \quart{3}{7}{6}{100} \\
    & 3 &      Java, novice &    6 &  7 & \quart{2}{7}{6}{100} \\
    &  3 &      C++, expert &    7 &  8 & \quart{3}{8}{7}{100} \\
\end{tabular}}}
\label{tbl:b5}
\end{table*}

\tbl{b5} presents our results on production rate and defects in 4 groups. From this Table, we make the following observations.

\bi

\item Despite numerous studies in the past that endorsed this effect, groups 1 and 2 reveal no effect on developers with years of prolonged programming experience. In other words, \textit{Novice} developers were as productive and induced the same amount of defects as \textit{expert} developers. 

\item Our earlier results confirm some programming languages to have an effect on ``production rate'' and ``quality'' (defects). Thus to check whether ``years of experience'' also influence developers using different programming languages, we segregate the \textit{expert} and \textit{novice} population by programming languages to find the following:

\bi
\item ``Years of experience'' has less influence on ``defects'' among developers using different programming languages. 
Like in our earlier results seen in \tbl{b2}, overall C\# and VB \textit{novices} portray better quality (fewer defects) than developers of three other languages. This also implies that strangely C\# and VB \textit{novices} portray better quality (fewer defects) than \textit{experts}.

\ei 
\ei

\textit{Apprentice Law} is only supported on the lines of production rate, only for Java and C++ developers (2 of 5 groups of developers), and has no influence in mitigating defects. Our evidence supports the counterclaim that practical industrial experience has little to do with expertise. There is no noticeable performance difference among \textit{experts} and \textit{novices} (Groups 1 and 2). We believe the conditions for deliberate practice~\cite{ericsson2004deliberate} are not achieved in normal work; thus, \textit{years of experience} has limited benefit.

Hence, overall, we say that. 

\begin{tcolorbox}
Belief 5: \textit{Experienced developers did not necessarily write better (fewer defects) programs on time.}
\end{tcolorbox}

\section{Aggregated Analysis \& Discussion}\label{tion:discussion}

\begin{figure}[h]
\begin{center}
\frame{\includegraphics[width=4.75in,keepaspectratio]{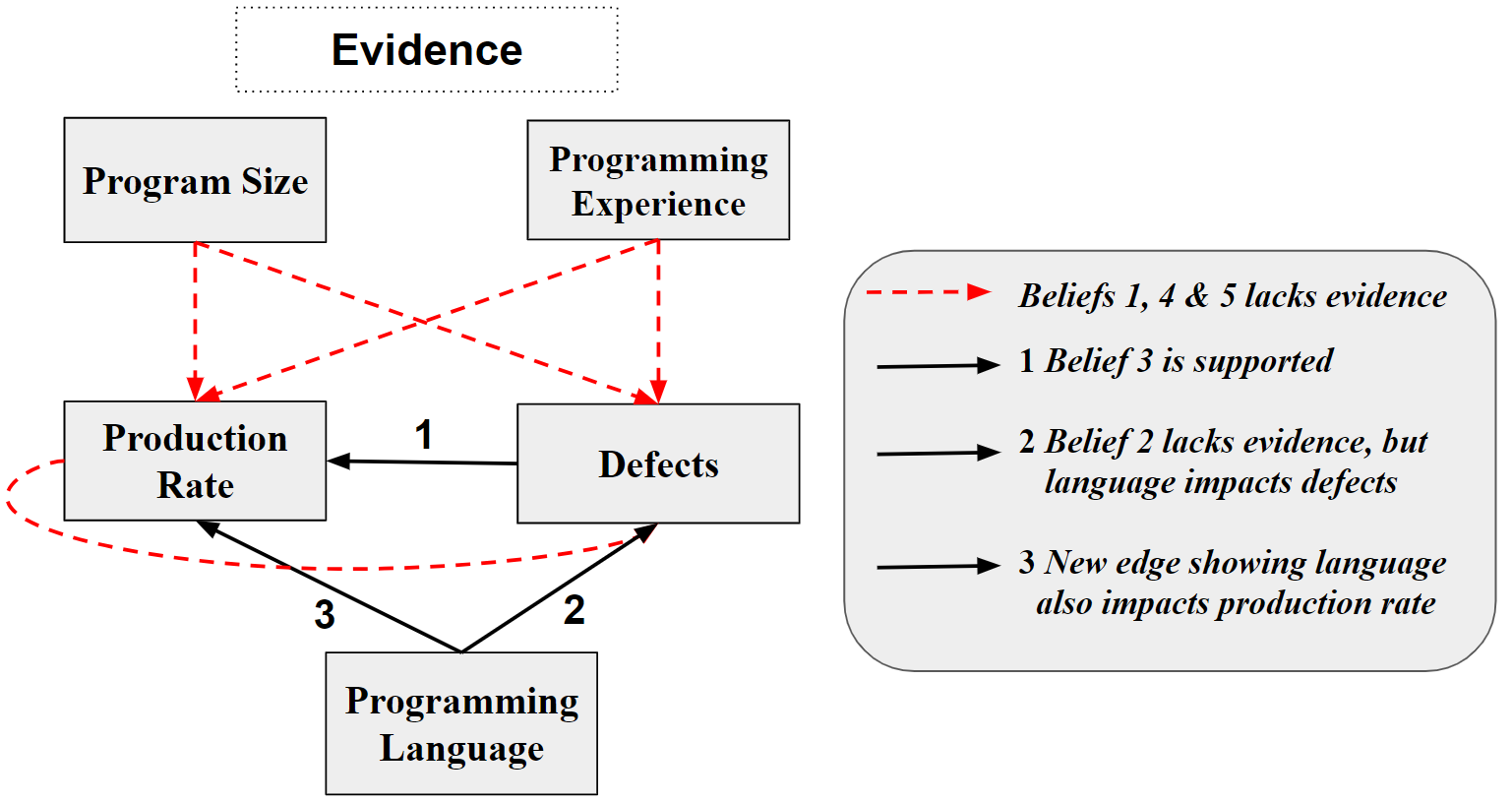}}
\caption{A copy of \fig{triangulate_question} but, edges are redrawn based on the results obtained from assessing the five beliefs in \tion{assessing}. Here a line edge (black) between two entities denote an effect backed by empirical evidence. 
} 
\label{fig:triangulate}
\end{center}
\end{figure}

Earlier in \fig{triangulate_question}, we presented the relationship between beliefs and their entities as recorded in literature. We revisit that figure using the evidence from assessing the five beliefs in \tion{assessing} and re-drew that graph in \fig{triangulate}. We did this to infer a combined opinion by aggregating results from multiple experiments similar to the notion of meta-analysis in statistics. Thus using \fig{triangulate}, we address the three RQ's we had asked earlier.

\subsection{\textbf{RQ1: Why beliefs diverge among practitioners?}}

Given that the five beliefs we chose are decades-old prevalent beliefs, we naturally expected strong support, but surprisingly, our analysis showed none of these beliefs is \textit{strongly} supported \textit{presently}. It is important to note that such beliefs naturally hold in practice  ~\cite{endres2003handbook,passos11,devanbu2016belief}, and this is not to say that these beliefs \textit{were} not true. 

We reason below that a probable source of divergence of beliefs among practitioners~\cite{devanbu2016belief} could arise from misinterpreting effects by observing partial evidence. For example, recall the effect reported in belief 2 that, 

\begin{center}
    {\em Programs written using non-OO languages naturally induce more defects.}
\end{center}

Although the results of belief 2 from \tion{b2} show that programs written in OO languages C\# and VB showed better quality (fewer defects), we did not endorse this effect because programs written in the two other OO programming languages C++ and Java shared defects similar or higher to those written in C (non-OO). 
While we \underline{{\em can not}}   say why defects were lower in C\# and VB, but we \underline{{\em can}} say that it is not due to the OO paradigm. To that end, it is reasonable to imagine that practitioners with a narrow scope who work only with C\# and C-based projects will hold on to this belief. Note that similar examples of misinterpretations backed by partial evidence can be weaved from the other beliefs we do not endorse. Thus we conjecture that practitioners could believe some effect to hold in their work due to the lack of a broader perspective.

Lastly, looking at \fig{triangulate} amidst overall negative results, we found support for belief 3 titled ``\textit{Quality entails productivity}'' which is unaffected among 4 of 5 programming languages we analyzed, which we endorse.

Accordingly, we say:
\begin{tcolorbox}
{\textit{Apart from belief 3 titled ``\textit{Quality entails productivity}'', none of the other beliefs are supported.}}
\end{tcolorbox}

\subsection{\textbf{RQ2: What is the relationship between Productivity, Quality, and Expertise?}}

If studies like this can confirm associations among these three entities (Productivity, Quality, and Expertise), practitioners can make better choices during their project life-cycle. Associations such as
\bi
\item Experienced developers produce a quality deliverable on time.
\item Early quality assurance activities can ensure faster delivery.
\ei

Using the directed edges of the graph shown in \fig{triangulate} we find:

\bi

\item Belief 5 results do not reveal any beneficial effect of \textit{years of developer experience} on software quality and only make some group of developers (specifically Java and C++) more productive.  
\item On the other hand, belief 3 results confirm that early quality assurance facilitates on-time delivery. That association is unaffected among different programming languages or task complexity.

\ei

Accordingly, we say:
\begin{tcolorbox}
{\textit{A focus on quality early in the project life-cycle minimizes rework, but programming experience neither improves production rate nor mitigates defects.}}
\end{tcolorbox}

Notably, it is now apparent from the result of the beliefs and these discussions that some programming languages were better than others. We discuss that next in RQ3.

\subsection{\textbf{RQ3: What impacts Productivity and Quality? }}

We advocated the strength of all the beliefs in this study, either using production-rate, quality (defects), or both. Analogously practitioners, especially managers during their project life-cycle, monitor and report the project's health using production rate and quality. There is no value in merely recording these two measures, but it is fruitful to understand what (factors) control them. This would help managers execute actionable steps to on-time defect-free delivery—factors such as developer expertise, programming language, tools, etc. If expertise is important, then practitioners may invest more budget to onboard expert developers. The belief results in \tion{assessing}  show that developers completing tasks using some programming languages were better (more productive and induced fewer defects) than others. Thus we think it is useful to analyze the results of beliefs holistically in the context of programming language and developer expertise  as follows:

\bi

\item Results of beliefs 4 and 5 confirm production rate is better among C\# developers, irrespective of their programming expertise. 

\item Years of developer experience only made an impact on the production rate of  C++ and Java developers. But note, Java \textit{expert}'s production rate is significantly lower than C\# \textit{novices}.

\item Results of belief 5 strangely confirm defects are lower among C\# and VB \textit{novices} than \textit{experts}.

\item Results of belief 1 shows C\# developers wrote longer programs similar to C developers, but C\# developers were more productive (higher production rate) and induced fewer defects than C developers.
\ei

\subsubsection{Why?}
Note that usability studies that discuss the learning curve of various programming languages are still largely under-explored. Such studies are very much required to reason for our results. Still, we conjecture as to why specific results are happening as follows:

\bi
\item Although C\# is derived from C and C++; it mostly resembles Java~\cite{eaddy2001c}. Additionally, C\# has many improvements over Java and C++. One interesting improvement to this discussion is C\#'s simplified syntax in comparison to Java. We conjecture that C\#'s less verbosity over Java and C++ makes it easier to learn, especially for \textit{novices}. Less verbosity implies faster coding in less time, therefore a better production rate~\cite{gupta2004good}. 

\item We note that C\# developers have written programs of length (LOC) similar to that of C. But note C\# programs would cover more functionality than C programs of similar length. It is because C\# has in-built functions (like string-operations, math-utilities, etc.) available as part of the library, whereas C developers have to code such functions from scratch~\cite{hejlsberg2006c}. Coding such functions in C may induce defects. Hence, we presume C\# developers have completed programs with better production-rate and lower defects than C developers.

\item Lastly, to understand why programs written by C\# and VB \textit{novices} had fewer defects than \textit{experts}, requires a deeper investigation and beyond the scope of this work. Having said that, such controversial results of \textit{novices} surpassing \textit{experts} are reported in the past~\cite{adelson1984novices,prumper1992some}. 

\ei

Apart from assessing belief 3, the results of the remaining four beliefs indicated the influence of developers using some programming languages over another.  Overall, developers were productive and introduced lesser defects when using modern languages (C\#, and VB). 

Accordingly, we say:

\begin{tcolorbox}
{Programming languages --- \textit{C\# and VB developers wrote programs with fewer defects. Specifically, C\# developers were most productive (among four other groups of developers who wrote programs in C, C++, Java, and VB).}}
\end{tcolorbox}

\section{Threats to validity}\label{tion:threats}
We draw the following subsections from Wohlin et al. ~\cite{wohlin2012experimentation} (first conceived by Cook and Campbell ~\cite{campbell1979quasi})
\subsection{Conclusion Validity}

Construct validity checks whether the findings could be incorrect because of the operationalizations of the concepts, incorrect modeling, or misleading data. 

 While Johnson and Disney found that 5\% of the data was incorrect~\cite{Johnson1999}, their study used manual recording, transcription, and computation.  Only the data prior to 1996 in this study were manually recorded. The vast majority used an Excel spreadsheet for initial data entry and all derived computations. Authorized PSP instructors also verified the data. A concern may arise due to the nature of the data set composed of only ten tasks (assignments). But, using similar data set, certain useful observations have been made in the past, and those are reported in these SE articles ~\cite{menzies2017delayed,nichols2019end}.  Another concern may arise from the sample size of our data set and the fact that OS can offer much more data at scale. But to directly check whether our conclusions apply to practice, a data corpus of industry nature is needed. More than 90\% of the tasks that happened in the industry across various geographies, and we only report statistically significant results, as mentioned in \tion{threshold}. 

Lastly, some studies operationalize quality as defect density. But defect density could be sensitive to the verbosity of either the programming language or the programmer. That is, the same assignments could contain the same number of defects yet differ because one program has more lines of code than the other. Defect density is often used because different programs cannot be directly compared. Because this data replicates the same task across multiple developers,  quality is best measured by the total number of defects in code, design, and test accordingly.

\subsection{Internal Validity} \label{tion:internal}

Threats to internal validity concern the causal relationship due to the artifacts of the study design and execution.  It may also include factors that have not been controlled or measured or study execution introducing some unintended factors. 

PSP course's emphasis on measuring production, estimation, and quality could have influenced the developer’s performance.  The mitigation was that the developers were not in any sort of competition with each other; instead, they were instructed to take consistent data to measure their performance trend. Also, there are no overlaps, i.e., the same developer completed the ten tasks only once using a programming language. Other factors that were uncontrolled include experience with a specific programming language or aspects of the development environment in which the class was taken.

Analyst bias in conducting the research is always a potential threat.
This is minimized because the data was collected for an entirely different purpose over an extended time by several independent individuals.  We further minimize this threat by relying on quantitative data and fully revealing that data.  While the tasks are unique, the underlying skills are somewhat consistent. The problems require reading input, writing output, performing basic data manipulation with lists, sorts, modular decomposition into subroutines, employing iterations, and conditional logic. There is some difference in that the two exercises place more emphasis on text manipulation, and a couple of others require nested floating-point iteration structures.

Lastly, we do not consider PSP as a treatment, but we use that data to observe evidence in the prevalent beliefs we evaluated. We do not question the authenticity of these beliefs in the past, given the notable increase in the number of programming languages, supporting tools, memory, computation power, and online workforce. We question the relevancy of these beliefs presently in \tion{assessing}.

\subsection{External validity}

The domain of the programs is not representative of all software development. The tasks were principally numeric and statistical. Nonetheless, they included the standard elements of modular design,  input, output, and control structures common to many professional programs.  The numeric specifications were provided; therefore, no special domain knowledge was required.  Production rates and defect rates will likely differ across specific domains. The programs were not intended to be of production quality; therefore, the test cases were not extensive. The goal of this work was to show that it is both possible and important to revisit old beliefs (and to advise practitioners to regularly monitor and discard effects that are not backed by evidence). The above results show that this is indeed possible.

As to specifics of our conclusions,
the set of programming languages we explore may not cover some of the recent trending web development languages like PHP, Ruby, etc. Thus we do not claim our results to generalize to all projects. On the other hand, 
 the languages we analyzed are in existence for decades in long-living proprietary software systems (in banks, healthcare, etc.) and will remain prominent in the future (if only for maintenance reasons).

\section{Summary}
\label{tion:conclusion}

Through extensive evaluation of five old SE  beliefs (originated between 1969 - 1993) in a controlled environment, we find support for one belief titled ``\textit{Quality entails productivity}.'' That implies on-time delivery is achieved with a quality-driven focus. Four other beliefs we assessed are not supported; uncertainties in the results of those beliefs portrayed how practitioners with a narrow scope could misinterpret specific effects to hold in their work.

Notably, we observed programming language to be a better indicator of software quality and production rate than years of developer experience. In other words, production rate and software quality varied for different programming languages. Overall, irrespective of the programming experience, C\# developers delivered tasks on time (productive)  with fewer defects. Prolonged programming experience only influenced Java and C++ developers to be more productive,  but it did not make them better (``quality'' and ``productivity'') than C\# developers.

In the future, a natural extension of this work is to check other prevalent beliefs that comment on SE phases like requirements and integration. Further, we would also like to identify additional factors like programming language that cause divergence of beliefs among practitioners.

\section{Implications for Practice}\label{tion:ifp}

Our results reinforce the recent findings of Shrikanth \& Menzies ~\cite{shrikanth2020assessing} and others in the past ~\cite{passos11, devanbu2016belief}, which is practitioners should not inherently believe their past will hold in the present. Like peer SE researchers ~\cite{monden2017examining, shull}, we suggest all practitioners, especially subject matter experts, consider assessing a handful of beliefs empirically from time to time to understand what works for their organization. Specifically, our current results prescribe,

\bi

\item Practitioners should emphasize quality right from the early stages of their projects.

\item Practitioners should be less concerned about programming experience and more concerned about programming language. 

\ei

\section*{Acknowledgements}

This work was partially supported by NSF grant \#1908762. Personal Software Process\textsuperscript{SM} and PSP\textsuperscript{SM} are service marks of Carnegie Mellon University.

\clearpage 
\bibliographystyle{plain}
\bibliography{bib_references}

\clearpage 
\section*{}
\begin{wrapfigure}{l}{25mm} 
    \includegraphics[width=1in,clip,keepaspectratio]{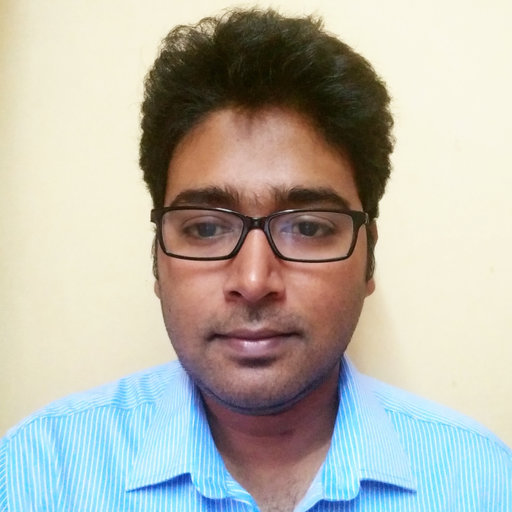}
  \end{wrapfigure}\par
  \textbf{N.C. Shrikanth} is a Ph.D. student in Computer Science at North Carolina State University. He practiced software engineering in India for nine years with three organizations (Accenture Labs, ABB India Ltd, and Infosys Ltd). His research interest includes software engineering and machine learning with a focus on software quality.  \url{https://snaraya7.github.io/}
\\\\\\
\begin{wrapfigure}{l}{25mm} 
    \includegraphics[width=1in,clip,keepaspectratio]{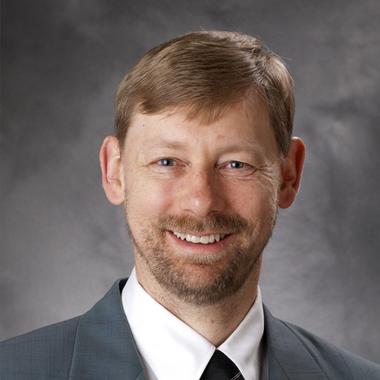}
  \end{wrapfigure}\par
  \textbf{William Nichols} is a senior member of the Software Engineering Institute’s technical staff at Carnegie Mellon University. He has more than 25 years of technical and management experience in developing scientific, engineering, and business systems. During his tenure at the SEI, Dr. Nichols has worked extensively with the Team Software Process (TSP) Initiative, where he currently serves as a Personal Software Process (PSP)
instructor and a TSP Mentor Coach. Current research interests include modeling the software development process and the development of secure software systems. Dr. Nichols is a Senior Member of IEEE and a member of ACM.
\\
\section*{}
\begin{wrapfigure}{l}{25mm} 
    \includegraphics[width=1in,clip,keepaspectratio]{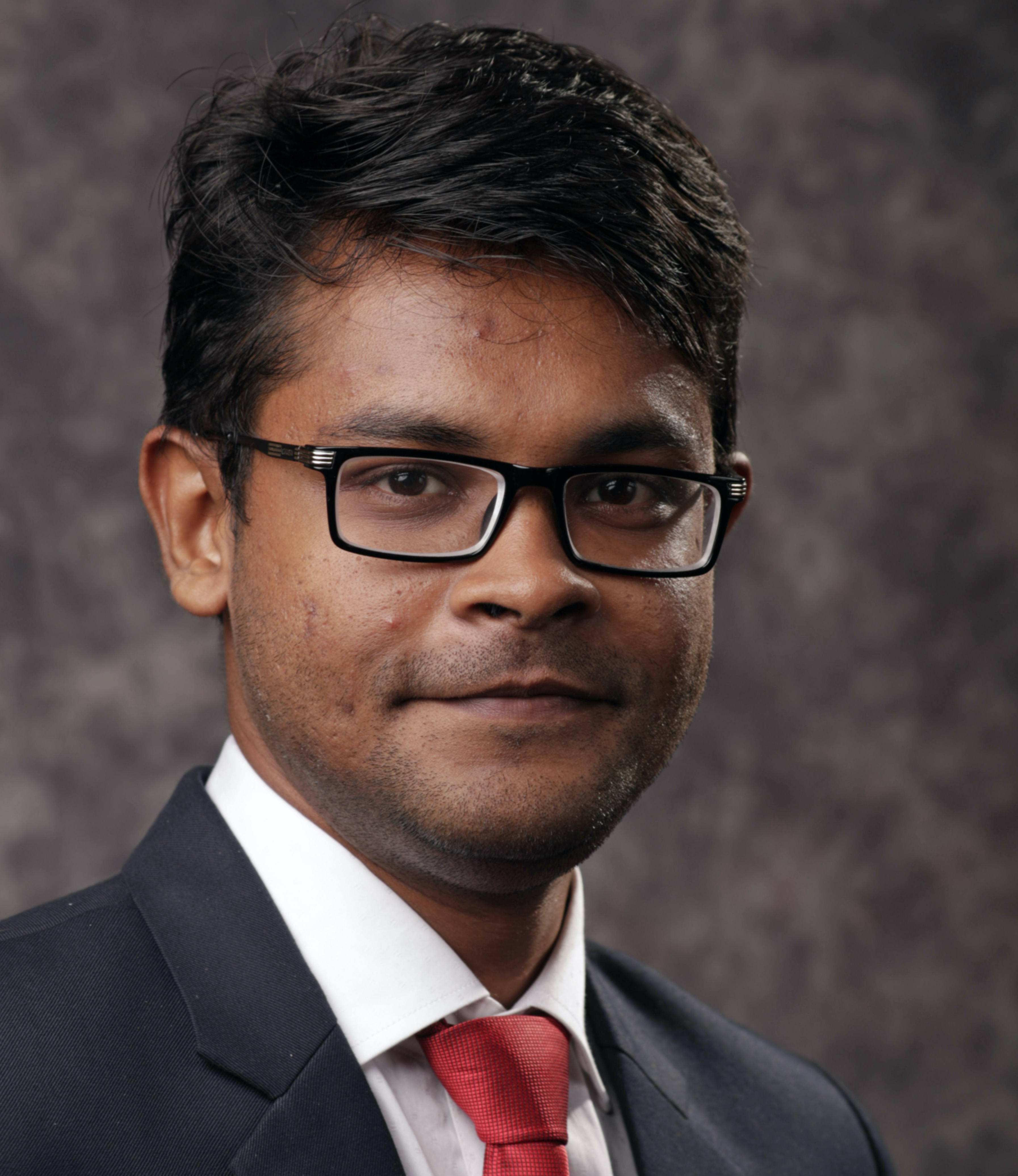}
  \end{wrapfigure}\par
  \textbf{Fahmid Morshed Fahid} is a third-year Ph.D. student in Computer Science at North Carolina State University. Before joining the Ph.D. program, he worked for two years in the software industry as a software engineer in Bangladesh (Reve Systems Ltd). His interests include data mining, AI in software engineering, reinforcement learning, and intelligent tutoring systems. 
  \url{https://fahmidmorshed.github.io}
\\\\
\section*{}
\begin{wrapfigure}{l}{25mm} 
    \includegraphics[width=1in,clip,keepaspectratio]{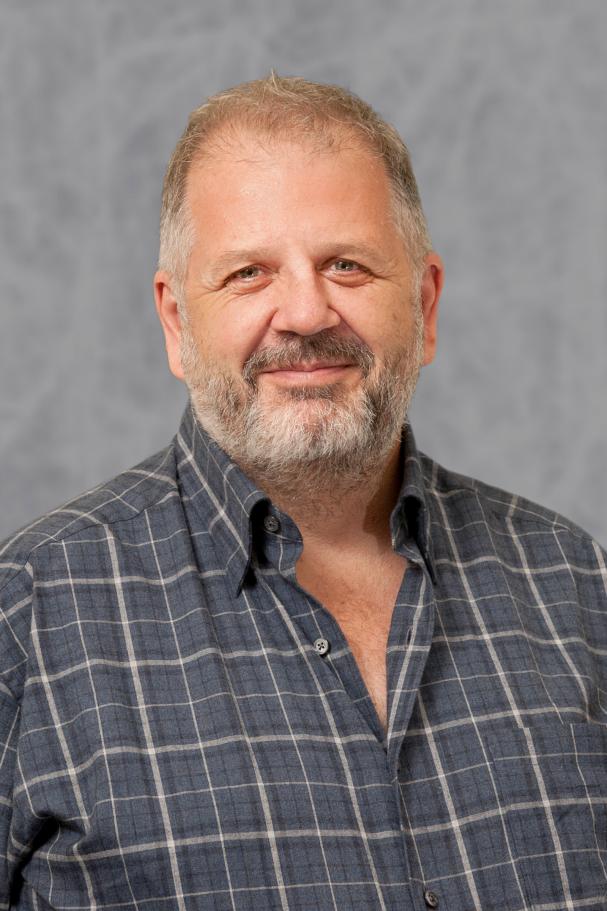}
  \end{wrapfigure}\par
  \textbf{Dr. Tim Menzies} is a Professor in CS at  North Carolina State University.  His research interests include software engineering (SE), data mining, artificial intelligence, search-based SE, and open access science. \url{http://menzies.us}

\end{document}